\documentclass[sigconf]{acmart} 
\setcopyright{none}
\AtBeginDocument{%
  \providecommand\BibTeX{{%
    \normalfont B\kern-0.5em{\scshape i\kern-0.25em b}\kern-0.8em\TeX}}}

\setcopyright{none}

\begin{document}


\title{Beyond Our Behavior:\\ The GDPR and Humanistic Personalized Recommendation}

\author{Travis Greene \& Galit Shmueli}
\email{travis.greene@iss.nthu.edu.tw, galit.shmueli@iss.nthu.edu.tw}
\affiliation{%
  \institution{Institute of Service Science, National Tsing Hua University}
  \city{Hsinchu}
  \state{Taiwan}
}
\renewcommand{\shortauthors}{Greene and Shmueli}
\begin{abstract}
   Personalization should take the human person seriously. This requires a deeper understanding of how recommender systems can shape both our self-understanding and identity. We unpack key European humanistic and philosophical ideas underlying the General Data Protection Regulation (GDPR) and propose a new paradigm of humanistic personalization. Humanistic personalization responds to the IEEE's call for Ethically Aligned Design (EAD) and is based on fundamental human capacities and values. Humanistic personalization focuses on narrative accuracy: the subjective fit between a person's self-narrative and both the input (personal data) and output of a recommender system. In doing so, we re-frame the distinction between implicit and explicit data collection as one of nonconscious (``organismic'') behavior and conscious (``reflective'') action. This distinction raises important ethical and interpretive issues related to agency, self-understanding, and political participation. Finally, we discuss how an emphasis on narrative accuracy can reduce opportunities for epistemic injustice done to data subjects. 
\end{abstract}

\settopmatter{printfolios=true}
\maketitle
\section{Introduction}
Machine learning-backed personalized services, among them recommender systems (RS), have become a permanent fixture in our increasingly digital lives. Personalization relies on vast quantities of behavioral big data (BBD): the personal data generated when humans interact with apps, devices, and social networks \cite{shmueli2017analyzing}. Besides supplying bases for recommendations and ``hypernudges'' \cite{yeung2017hypernudge} towards these recommendations, BBD are the essential raw material of our digital representations. But to what extent do these logged digital behaviors truly reflect who we are as persons? 

As life steadily moves online, the digital representations of persons take on legal and moral importance \cite{van2017decisional}. Influential European legal theorists and philosophers have even written an \emph{Onlife Manifesto}, shaping the discourse around what it means to be human in the digital age \cite{floridi2015onlife}. At the same time, the IEEE has articulated a vision of Ethically Aligned Design (EAD) that empowers ``individuals to curate their identities and manage the ethical implications of their data'' \cite{ieee2018}. 

The current domain of RS is largely focused on problems of ``practical identity,'' which include various methods for constructing, identifying, and linking various data representations of users \cite{de2010identity, manders2010practical}. In contrast, philosophers and social scientists are concerned with issues of moral, social, and increasingly, the narrative identities of persons. This separation of spheres of activity is no longer acceptable given the potential effects RS can have on the person.  We thus propose \emph{humanistic personalization} (HP) as a novel guide for thinking about and designing personalized RS. The HP paradigm of mutual, dialogic participation obliges RS designers to promote and respect the unique capacities of persons to create and modify personal narratives over time. 

Our paper builds on but differs from recent work on RS value-alignment \cite{strayyou}, self-actualization  \cite{knijnenburg2016recommender}, qualitative evaluation metrics \cite{mcnee2006being}, and the epistemic and ethical problems of over-reliance on behavioral data \cite{ekstrand2016behaviorism}.  

Most notably, we use the General Data Protection Regulation (GDPR) as the basis for ethical and legal norms instead of adding to the current morass of competing principles, guidelines, and frameworks for Ethical AI/ML (see e.g., \cite{mittelstadt2019principles}). Tying ethical principles to legal norms via the GDPR is valuable because, in principle, the GDPR applies to any data controller processing the personal data of data subjects residing in the EU \cite{greene2019adjusting}. Further, law is institutionally-backed to achieve greater behavioral compliance via the state's monopoly on the legitimate use of physical force \cite{cane2002responsibility, hart2012concept}. Thus our conception of HP is likely to foster greater compliance by industry and researchers around the globe, even if some minor details of the GDPR have not yet been fully worked-out. In sum, we believe the GDPR provides a solid grounding for deep humanistic design principles relevant to RS, a view shared by a number of European legal scholars \cite{hijmans2018ethical, hildebrandt2015smart}.

Second, we re-frame \cite{ekstrand2016behaviorism}'s distinction between implicit/explicit feedback as one between conscious action and non-conscious behavior and perception. This has philosophical and moral implications for RS users.


Third, we cast doubt on the meaning of the structural regularity of said observed behavioral data, which are encoded and parameterized by the machine learning (ML) algorithms used to generate recommendations. We suggest these regularities may more reflect the structure of platform affordances rather than a person's underlying preferences. In this way, debates in the philosophy of sociology (e.g., the \emph{structure-agency} debate) may provide insights into areas where personalization is currently weak. A final and absolute determination of person's ``true'' preferences or interests may not be possible--a conclusion which dovetails with postmodern theory and complexity theory \cite{cilliers2006complexity}. The upshot of this diagnosis is that giving data subjects a constructive voice in the meaning and interpretation of digitally recorded events may be our best option. We introduce the legal basis supporting such a conclusion. 

\subsection{The Recommender Systems View}
Conventional collaborative filtering (CF)--the dominant paradigm in personalized recommendation--relies primarily on implicit data collection to drive recommendations \cite{covington2016deep}. CF leverages implicit data to determine a focal user's ``nearest neighbors'' \cite{burke2018balanced} and recommends items bought or consumed by the ``neighbors,'' but not yet by the focal user. In many large-scale production systems, implicit data take the form of browsing history, rating metadata, search queries, and social graphs. Implicit data collection relieves users of the task of explicitly rating items and can be easily automated and scaled \cite{nichols1998implicit}. Implicit data are a type of BBD. 

 During personalization, a user is represented roughly as follows. Database representations of a person--tabularized collections of logged behaviors in apps or on devices--are converted into feature vectors, permitting the computation of various proximity metrics between pairs of vectors.  Next, ``neighborhoods'' are derived from the set of neighbors in feature space deemed closest to a given user, according to a distance metric.  A 10-dimensional feature vector represents a person as an array of 10 numbers, obtained from measurements of observed behavior, thereby replacing the person with a single point in 10-dimensional feature space. To reduce computational costs, the dimensions of the original user-item matrix are often reduced into a lower ``latent space'' by techniques such as singular value decomposition (SVD). Recently, deep learning approaches to RS have used stacked denoising autoencoders to generate latent representations of users  \cite{vincent2010stacked}. Digital representations of persons thus take the form of feature sets of narrow, observed behaviors within an app or device and exclude their unique mental states, social and moral identities and personal narratives. 
 
 \subsection{A Humanist Critique of BBD-based RS}
 The behavioral focus of CF personalization exposes it to the same criticisms of behaviorism. Behaviorism rejects the study of \emph{causal} cognitive mechanisms of human behavior (beliefs, intentions, goals, values) in favor of measurable features of the behavioral environment \cite{skinner1965science}. The behaviorist model of the human person (abstracted to an ``organism'') is inherently at odds with that of the GDPR. BBD-based models are what a humanist might describe as a ``caricature'': a model not only overly simplistic and approximate, but which actively distorts reality \cite{gibbard1978economic}. 
 
 In the case of CFs, behaviors measured are \emph{self-interestedly} chosen by BBD platforms (e.g., Facebook or Google) and often driven by business--not scientific or ethical--goals, such as click through rate, engagement, or conversion. Further, the ability to manipulate digital environments at the individual level-- \emph{hypernudges}--obscures reliable interpretation of observed behaviors. This problem is particularly relevant for RS using \emph{online learning}, such as deep reinforcement learning \cite{zhao2019deep}.

Philosophers of science have long understood that measurement itself is a form of representation. As Bas Van Fraassen notes, ``a measurement outcome does not display what the measured entity is like, but what it 'looks like' in the measurement set-up'' \cite{van2010scientific}. The act of measuring something ``locates it in an ordered space of possible measurement outcomes,'' where this ordered space is built into the measuring instrument \cite{baird2004thing}. Likewise, observations produced by instrumentation (e.g., phones or apps) are inherently perspectival: they are sensitive only to a particular type of input and are never \emph{perfectly} transparent \cite{giere2010scientific}. Behavioral data are more precisely \emph{capta}, or the ``units that have been selected and harvested from the sum of all potential data'' \cite{kitchin2011code}. 

Feature sets of \emph{narrowly} defined and \emph{observed} (logged) behaviors within an app or device constitute our digital representations. Yet the \emph{meaning} of observed behaviors are under-determined since they could spring from a variety of possible mental states. By leaving out mental states, social and moral identity, and narrative aspects of human experience,  
BBD-based models of persons are \emph{representations} merely by ``stipulative fiat'' \cite{callender2006there} of RS designers, not because they capture \emph{essential} aspects of a person.

\subsection{Two Philosophical Views of Persons and Personal Data}
The act of inferring personal interests through digital behavior--the ostensible goal of personalization--leads to deep philosophical questions. For instance a) What \emph{kind} of objects are our data representations? and b) What is the relation between \emph{parts} (measurements) and \emph{wholes} (persons), especially when a \emph{global} objective function is optimized? 

Two influential accounts of how persons and personal data relate exist in the philosophy of information and marketing literature.  Yet we find both views to be legally and morally problematic. Our purpose in introducing these two points of view is not to attempt to hash out their metaphysical details, but instead to introduce readers to a new way of thinking about the complexities of technologically-mediated being. We describe each view next.

\subsubsection*{You (Really) are Your Data}
The Oxford philosopher of information Luciano \cite{floridi2005ontological} endorses what we term a ``realist'' view of personal data.  He maintains, ``You are your information,' so that anything done to your information is done to you, not to your belongings.'' To Floridi, personal data do not \emph{represent} you, but rather \emph{constitute} \emph{you} as \emph{you}: your personal identity and your personal data are functionally equivalent, though at different \emph{levels of abstraction}. 
Floridi contends that the GDPR's rights to informational privacy protect the constitution of one's identity. If data processors alter or delete personal data about you, they also alter or delete aspects of your personal identity. 

Floridi's view confers strong privacy safeguards for data subjects--Facebook's Emotional Contagion experiment \cite{kramer2014experimental} would be tantamount to psychological abuse. But Floridi's view falls short for other reasons. \emph{Persons} have rights, personal \emph{data} do not. \emph{Data} can be nearly infinitely cloned and copied, \emph{persons} cannot. Further, if Floridi is right, \emph{identity theft} would be akin to literal kidnapping \cite{shoemaker2010self}. Moreover, Floridi's view overstates the importance of how others classify us. Our social identities indeed depend on how others label us, but we are capable of \emph{choosing} to reject this label. Lastly, Floridi does not provide any account of \emph{why} certain kinds of personal data are more sensitive than others.

\subsubsection*{Dividuals: Persons or Posthumans?}
Postmodern marketing researchers and cultural theorists take a different approach. 
For them, databases and now RS construct \emph{dividuals}: digital entities derived from masses of individuals' personal data \cite{cheney2011new}. Dividuals are, in the words of poststructuralist philosopher Gilles Deleuze, ``cybernetic subject[s] made up of data points, codes, and passwords'' \cite{deleuze1992postscript}.  Matrix decomposition techniques, such as SVD, combine and re-assemble personal data to birth new marketable entities separate from the persons from whom they were derived. The resulting posthuman \emph{assemblages} are variously termed \emph{data doubles}, \emph{capta shadows} and \emph{digital personas} \cite{kitchin2011code}.  

This interpretation of digital identity is also problematic. Under the GDPR, only \emph{natural living persons have rights}--dividuals are outside the scope of the GDPR. Facebook's above-mentioned study would be legally and ethically kosher. Second, personalization does not divide up individual persons, but rather their digital representations. Unlike human individuals, nothing about these digital representations demands they remain unified wholes. Finally, recommendations must ultimately be linked to \emph{actual} living persons through unique identifiers, but it is unclear how dividuals relate to the persons from whom they were derived. While the dividual is an interesting metaphor, it has limited applicability in the real world. 

The remainder of this paper is structured as follows. First, we consider and critique the various epistemological and ethical blindspots of many current BBD-based RS based on ML algorithms (e.g., CF and reinforcement learning). Next, we lay out the legal and philosophical groundwork underlying the GDPR. We argue the conceptual seeds of humanistic personalization are rooted in the legal principles of a right to personality and in informational self-determination. We then connect these principles to the ideas of key Enlightenment and Romanticist thinkers, and derive some insights for RS design. Following that, we respond to the IEEE's call for EAD by introducing the notion of \emph{narrative accuracy} and its complement, epistemic injustice \cite{fricker2007epistemic}. The goal of narrative accuracy is to align predictions and explanations of RS with one's personal narrative. Finally, we consider potential dangers in emphasizing human subjectivity through narrative and discuss possible compromises. Ultimately, we hope this work contributes to the conceptual foundations of \emph{participatory design} of recommender systems \cite{balka2006inside, ekstrand2016behaviorism}, and \emph{third paradigm HCI} \cite{harrison2011making}.

\section{The Complexities of Human Behavior in the Digital Age}
We currently reside in what technologist Eli Pariser calls the ``uncanny valley of personalization'' \cite[pg. 65]{pariser2011filter}. Users know recommendations derive from our digital representations, but are often not quite sure how or why. Escaping this ``uncanny valley'' requires new, reflective perspectives on the design and impact of AI/ML. Our notion of HP envisions a shift from an implicit, behavior-based representation paradigm, dominated by our ``organismic'' interests, to one centered on conscious, explicit feedback \cite{ekstrand2016behaviorism} through the notion of dialogic narrative construction. 

Driving our discussion is a distinction we make between conscious action and nonconscious ``stimulus-response'' behavior.  The difference is essentially between ``what an agent does'' and  ``what merely happens to him'' \cite{frankfurt1978problem}. Philosophers have traditionally associated intentional action with conscious, rational beings and mere behavior with non-conscious states. Importantly for our arguments here, narrative identity presupposes conscious reflection and recounting of events. We realize there is a lot of grey area, but we believe our contentions to be supported by the emerging literature in cognitive psychology. 



Our main point in this section is simple. We assert that it is not that case that current RS personalization is so incredibly accurate, but that we, as linguistic, social, and physically-embodied animals, have deceived ourselves as to our potential for free movement and thought. In reality, much of what we do is conditioned by the perceived features of our social, physical, and now digital environments. This idea may seem obvious now, but it took philosophers centuries to overturn the Cartesian dualism between mind and body and realize the boundaries between subject and object were not clear and distinct. Our bodies, for instance, shape our perceptions of the world and our choice of linguistic metaphors \cite{lakoff2008metaphors}. 

On top of this, the increasing ubiquity of digital environments further limits and constrains what is humanly possible. Perhaps even worse, as degrees of freedom in digital environments are reduced, hermeneutic\footnote{Hermeneutics dealt with the interpretation of biblical texts, but was revitalized as a philosophical method of literary analysis in the 20th century.} problems of human action arise and present ethical problems. For one, the behaviorist assumptions behind the collection of implicit data fail to appreciate the important caveat that ``variable acts produce a constant result'' \cite{powers1973feedback}. When complex behaviors are broken down into overly-narrow ``sub-symbolic'' categories by BBD platforms and RS designers, intentions become decoupled from results. \footnote{We note that the English word ``category'' traces back to the Greek \emph{kategoreisthai}, which means ``publicly accuse.'' \cite[pg. 18]{bourdieu2015sociologie}} What is more, a clear one-to-one mapping of intentions to actions becomes impossible. One cannot intend to do what one cannot first identify. Misinterpretation appears baked into digital life and is worsened when automated systems can dynamically change digital environments in real-time.

\subsection{Ideomotor and Nonconscious Goal Directed Behavior}
More recently, neuroscientists and psychologists have studied how our brains and perceptual systems evolved to make quick and relatively accurate assessments of our social and natural environments, resulting in the nonconscious guidance of mental processes involved in interpersonal behavior and goal pursuit \cite[pg. 44]{hassin2004new, bargh2001automated}. So although the behaviorist paradigm in psychology is no longer dominant, it was only \emph{partly} misguided. Much of what we do is indeed determined by perceived environmental structures (e.g., affordances) and evolutionary drives, many of which operate nonconsciously. 

According to \cite{gauchou2012expression}, ideomotor actions are ``movements or behaviours that are unconsciously initiated, usually without an accompanying sense of conscious control.'' They are actions (in our sense, behaviors), typically triggered by environmental cues, that express information not consciously accessed. These researchers speculate that ``ideomotor actions, especially visually-guided ones, may reflect the operation of an `inner zombie'--a concurrent nonconscious system expressed primarily via motor action'' \cite{gauchou2012expression}. \cite{ferguson2008becoming} explains further, ``After people have been implicitly primed with cues closely linked in their memory with striving toward an end-point, they display behavior that meets classical criteria of motivation, such as persistence and resumption after an interruption.'' In other words, it can be impossible to tell from observation alone whether movements are conscious actions or nonconscious behavior. 

Yet new technology allows for the capture of increasingly fine-grained movements. As one example of how nonconscious behavior can potentially be leveraged by a RS, \cite{carrabis2014system} describes a patent for software correlating nonconscious elements of behavior with users' demographic characteristics. By analyzing the distinctive patterns of mouse trajectories, for instance, inferences about nonconscious activity can be drawn. The patent demonstrates how this could be used to personalize the presentation of website content:

\begin{quote}
 As individuals continue to become more accustomed to using digital devices, and activities hosted on digital devices, the concept of tracking an individual's nonconscious behaviors using the digital devices becomes increasingly convenient... [a] web server can prepare in real time what the next presentation and interface should be in order to capture more of the viewer's attention by presenting the web content in modalities which the viewer has nonconsciously selected. Thus, content is directed via a Smart web server to a viewer based on the viewer's nonconscious selection.
\end{quote}

On the flip side, if nonconscious behaviors are regular enough to be detected by ML techniques, then this suggests we might use ML to actively discount or re-weight behaviors in favor of conscious, intentional action. 

Moving now to the field of information retrieval, a major assumption is that the user is driven by an ``information need.'' Even when updated using Broder's ``trichotomy'' of web search types \cite{broder2002taxonomy} to include informational, navigational, and transactional searches, we still do not do justice to the complexities of human meaning and context. Such a model of human behavior abstracts a person to the level of its organismic interests. It is a caricature.

From a humanistic point of view, such research typically fails to distinguish between conscious and nonconscious ``interests'' and ``goals.'' As one example, \cite{guo2010ready} analyze mouse trajectories and scroll patterns for the ``intent to buy.'' Researchers also typically overlook key ethical implications of pre-defined ``intent'' categories, for example, that intentions are assumed to either fall under ``research'' or ``buy.'' For a philosopher, there is an important difference between behavior that is more or less evolutionarily hard-wired--essentially stimulus-response--and what is the result of deliberate, reasoned reflection on one's personal values. Law also recognizes a moral difference between the two. 

Humans are more than mere information-seeking organisms. A web query such as ``how can I get a six pack in 30 days'' and ``how can I help end human suffering'' are both informational search types driven by an assumed information need. Yet anyone but the most die-hard positivist would argue these observed queries (treated as bags of words) are qualitatively equivalent. Personalization should take the person seriously. If we recognize the human capacity for moral reflection, we have to treat the second search as inherently more meaningful and worthy of pursuit than the first. The question is, how might we design RS to promote this kind of reflective endorsement of our desires?

\subsection{Behavior Modification and Dark Patterns}
 Although controversial, behavior modification methods have been used by major BBD platforms to control digital environments and shape behavior towards ``guaranteed outcomes'' \cite[pg. 201]{zuboff2019age}. Derived from principles of behaviorist psychology, they are premised on the idea that ``once you understand the environmental events that cause behaviors to occur, you can change the events in the environment to alter behavior'' \cite[pg. 3]{miltenberger2011behavior}.  A subset of these techniques are the so-called ``dark patterns'' of design. Dark patterns are cases where designers use their knowledge of human behavior to implement deceptive functionality that is not in the user's best interest \cite{gray2018dark}. We note that no distinction is made between ``organismic'' and ``reflective'' interests here. 
 
 Many of these dark patterns stem from the ideas of influential design gurus, such as BJ Fogg and Nir Eyal, who are candid about the ways in which products\footnote{RS can be seen as metaproducts: products which recommend other products.}such as RS are ``persuasive technologies,'' based on ``choice architectures'' aimed at getting us ``hooked.'' These persuasive techniques include reduction, tailoring, tunneling, suggestion, self-monitoring, surveillance, and conditioning \cite{gray2018dark}. 

Some dark patterns are especially relevant to RS design, such as \emph{interface interference} and \emph{forced action} \cite{gray2018dark}. Typically, the ethics of dark patterns are seen from the designer's point of view. There the question is, ``Is it right to design this way?'' We suggest there is another equally interesting way to view persuasive technology and dark patterns--from the user's point of view. There the question is: ``If I didn't consciously choose this, does this `choice' reflect anything deep about who I am?''


\subsection{Problems with Context and Interpretation}
  Computer scientist Paul Dourish details how the positivist, Cartesian paradigm of clear and distinct separations of subject and object, knower and knowledge, dominate computer system design \cite{dourish2004action}. In the philosophically-informed version of HCI he advocates, ``our experience of the world is intimately tied to the ways in which we act in it'' \cite[pg. 18]{dourish2004action}. On this view, questions about the ``true'' boundaries of subject/object and system/environment are inherently irresolvable. Dourish also highlights a key aspect between what \cite{ekstrand2016behaviorism} call the intention/behavior gap. \cite[pg. 137]{dourish2004action} explains: 
  \begin{quote}
      When I click on the Buy now? button on a Web page, what matters to me is not that a database record has been updated...  but rather that, in a day or two, someone will turn up at my door carrying a copy of the book for me. I act through the computer system. In turn, this takes us to another element of the relationship between embodied interaction and intentionality.
  \end{quote}
  
  RS research that accounts for this problem of one event having two potential meanings is hard to find. In their work on context aware recommender systems (CARS), \cite{adomavicius2011context} paradoxically claim they rely on a ``representational'' approach and cite Dourish's work. Yet Dourish's paper is aimed at explaining the ways in which ``positivist/representational'' thinking behind ubiquitous computing is shortsighted and incomplete. He instead suggests drawing on ``phenomenological'' and ``embodied'' approaches to understanding context and its role in meaning creation. Dourish, however, does not consider legal norms in his discussion of how designers can be ``nudged'' to include these more nuanced, post-positivist conceptions of context into RS. We claim the GDPR provides grounds for doing so, irrespective of the underlying economic or business logic.

\subsection{Behavior as Text}
 The philosopher Paul Ricoeur connects language and behavior in a unique way relevant to BBD and RS, and social science more generally. He holds that we can view behavior as \emph{text} and thus apply hermeneutic methods to its interpretation, much as we might to a book or speech \cite{ricoeur1971model}. He calls this aspect of text \emph{distanciation}. Distanciation breaks the link between writing and its original context (and its intended audience), granting the text semantic autonomy while also creating new problems of reference. Ricoeur alleges that to understand a text, the reader must ``unfold'' the possibility of being indicated by the text, an act he refers to as \emph{appropriation} \cite[pg. 55]{thompson1981critical}. In other words, interpretation involves self-conscious reflection on both the proposed meaning of the text, and how such a meaning might contribute to one's own self-understanding in relation to others (its author). Ricoeur claims the creativity of literary narrative and self-identity are driven by this process of ``imaginative variations'' on an underlying invariant structure of text or character \cite[pg. 148]{ricoeur1994oneself}.

Once we understand digital behavior as text, we can go one step further and apply the insights of Jacques Derrida towards the goal of BBD interpretation. Derrida asserts that speech requires presence, which makes the context of the utterance clear \cite{derrida1988limited}. Traditionally, you had to be near someone to hear what came from their mouth. This nearness provided the context for the utterance: in case of any confusion, its referent could be pointed to or gestured at by the speaker. Meaning could be dialogically negotiated relatively easily. But the power of writing is that it detaches the written sign (the word/signifier/symbol) from the writer in its original, intended context (the signified). Derrida claimed that writing is the death of the author because it still functions as a sign in his absence  \cite[pg. xli]{derrida2016grammatology}. We can interpret the written mark in any conceivable way once we have broken it from its original context of production. Derrida's point is this move from presence (speech) to absence (writing) is both problematic and creative. Re-presentation is an attempt to recreate the presence of the author of the text in some new, different context.

The point in introducing Derrida is to show that once detached from the presence of data subjects, personal data become ``public signs'' which can be algorithmically manipulated and combined in novel ways by RS designers. The downside is that we can now never know the original ``true'' intention in the absence of the original act of data production. We must be satisfied to re-present and not present the author of the data. We thus face a crisis of interpretation. What to do? If we follow the GDPR, we let the data subjects themselves decide. 

\subsection{The Agency-Structure Debate and Reinforcement Learning}
Sociologist Anthony Giddens introduces the idea of \emph{structuration} to analyze how societies evolve in terms of systems, structures, and rules. Systems are the relations between actors, organized as regular social practices, while structure is the unseen rules and resources deriving from the system; finally, rules are procedures for social interaction which constitute meaning and sanction various conduct \cite[pg. 59]{manicas2006realist}. Structuration emphasizes the recursive duality of social structures--they have both synchronic and diachronic aspects--implying that by performing behaviors and following rules, agents simultaneously change and cement existing social structures. 

In Giddens' view, it is mistaken to believe there were agents before societies, as was assumed by Locke or Hobbes. Instead, Giddens argues all persons are born into existing societies, and social structures pre-exist for them as the ongoing activities of members. Agents \emph{become} persons in a society, and by their actions within it, reproduce and transform it. This view suggests two opposing conclusions: the actions of agents in society are to some extent determined by existing structures and rules in society, yet by acting each actor transforms society and gives it a unique history. The philosophical question is, how much of what we do in society is determined by its structure and how much derives from our capacity as free agents?

A similar question arises in our interactions with digital environments. As an analogy, we can think of reinforcement learning (RL) used in RS. The goal of such an RL system is to find an optimal recommendation policy as it continually explores and exploits its state space \cite{garcia2015comprehensive}. This adaptive process is what drives RL personalization and its various applications in search, ranking, and web-page/app optimization. In the contextual bandit approach, for instance, exploitation translates to recommending items predicted to best match users' preferences, while exploration translates to giving random recommendations hoping to receive more ``feedback'' \cite{zhao2019deep}. 

As one prominent example, Facebook uses its open source Horizon platform to send ``personalized'' push notification and updates to millions of users. \cite{gauci2018horizon} explains how Facebook uses a Markov Decision Process (MDP) to serve personalized recommendations:
\begin{quote}
    The Markov Decision Process is based on a sequence of notification candidates for a particular person. The actions here are sending and
dropping the notification, and the state describes a set of features about the person and the notification candidate. There are rewards for interactions and activity on Facebook, with a penalty for sending the notification to control the volume of notifications sent. The policy optimizes for the long term
value and is able to capture incremental effects of sending the notification by comparing the Q-values of the send and don't send action.
\end{quote}

Yet this approach poses many ethical questions, besides the obvious ones related to using a set of predefined user features. The other issue is the extent to which nonconscious goal-directed behavior may be driving many of the observed trajectories given a specific policy. But there are deeper problems of interpretation due to the positivist formalization of RL, in which it is assumed the truth is absolute, a collection of brute facts seen from a view from nowhere (see, e.g., \cite{dourish2004we}).  

First, the interpretations of states (e.g., the sets of categories describing the person and the notification) in the agent's state space are pre-defined by the system designers. Second, the interpretations of actions (e.g., sending and dropping notifications) in the action space are defined by system designers. Third, rewards are chosen by system designers and typically try to maximize things like click through rate, dwell time, or business revenue, which may have little to no value from the data subject's perspective. Fourth, any constraints on the set of allowable policies from within the policy space are set by system designers. Taken together, these problems cast doubt on aspirations of personalization by RL systems, if we are to take the person and his epistemic capacities seriously. Few if any of the design parameters capture what the user would reflectively endorse as the interpretation of an action or state, or the selection of the reward to be maximized. These tensions suggest an area ripe for research on how a person's moral values can influence choices such as the reward and criteria for allowable policies. We also see promise in connecting ``teacher advice'' for safe exploration \cite{garcia2015comprehensive} by RL with persons' moral values and personal identities.

\section{The GDPR View}
This section lays out the basis for our reading of the GDPR, which we believe supports a new focus on narrative accuracy as an orienting ideal for RS. By the end, we hope to justify the legal scholar Mireille Hildebrandt's definition of privacy (in the European context) as, ``the freedom from unreasonable constraints on the construction of one's own identity'' \cite[pg. 80]{hildebrandt2015smart}.

Spurred by revelations of US government surveillance programs, the ubiquity of AI/ML in industry, and the ``datafication'' of society, European policymakers aimed to re-conceive the role of digital technology in society \cite{greene2019adjusting}. Besides creating a ``Single Digital Market'' across Europe, lawmakers also hoped to strengthen the rights of individuals to protect and control their personal data.
The distinctive nature of the European experience and philosophical tradition is reflected in the GDPR and distinguishes it from other recent regulations in California (CCPA) and China (Cybersecurity Law, CSL). 

The result was the European Union's 2018 General Data Protection Regulation (GDPR), which updated the 1995 Directive and built on the legal foundations of the 2000 Charter of Fundamental Rights of the European Union (CFR). The CFR recognizes fundamental rights to privacy and protection of personal data for all persons in the ``human community.''  We note that one reason for updating the 1995 Directive to the GDPR was to maintain consistency with the CFR's rights to data protection and privacy \cite{zarsky2016incompatible}. 
 Accordingly, much of the content of the GDPR was included, not because of anything related to technological advance or business concerns, but because it was needed for legal coherence with broader European ideas about human rights and privacy. 

The GDPR addresses the storage, collection, and processing of personal data. The GDPR also regulates the automated processing and algorithmic profiling of personal data. These terms were left intentionally vague as to encompass future developments in technology \cite{rosen2011right}. In the language of the GDPR, \emph{data subjects} are ``natural living persons''; \emph{personal data} are ``any information relating to an identifiable natural person''; and \emph{algorithmic profiling} is any kind of ``automated processing of personal data used to predict a natural person's interests or preferences'' (among many other predictive targets) with ``significant'' (legal or otherwise) effects on the natural living person. Personalization, and therefore RS, falls under algorithmic profiling. 

Unlike the US, where personal data processing occurs largely via an ``opt-out'' mechanism, the European approach is based on data subjects ``opting-in'' to processing \cite{Weiss2016}. In other words, EU data subjects must decide to  give consent to personal data processing when other legal bases of processing are not present. When no legal bases are present and the data subject has not consented to the processing of his personal data, processing cannot take place.\footnote{Article 6 lays out the six lawful bases of personal data processing}

Under the GDPR, data subjects inherit many of the same rights they had under the 1995 Directive, plus a few notable additions. Articles 12-23 spell out these rights. Data subjects now enjoy the \emph{right to be forgotten} (data subjects can request deletion of their data) and the \emph{right to data portability} (data subjects can request a portable, electronic copy of their data). 

Generally speaking, the rights of data subjects under the GDPR can be categorized as related to transparency (i.e., clear and unambiguous consent, communication with data subjects should be clear and easily intelligible), information and access (i.e., who collected the data and for what purpose(s)), rectification and erasure (i.e., allowing data subjects to correct false information and delete old information), and objection to (automated) processing (i.e., removing consent to processing of any personal data including algorithmic decision-making).

The GDPR places emphasis on the broad social effects of personal data processing. Recital 4 states that the purpose of personal data processing is to ``serve mankind.'' Consequently, an individual's rights to object to processing are not absolute; they must be weighed against broader societal benefits of such processing. The process of weighing is called the \emph{principle of proportionality}. An example of such a right is the new right to be forgotten, which, in some cases may conflict with the right to information access, and other retrieval and archiving goals, especially when dealing with publicly accountable figures, such as politicians \cite{andrade2012oblivion}. In this sense, the GDPR can be seen as taking somewhat of a utilitarian approach to personal data processing. 

\subsection{Informational Self-determination and the Right to Personality}
Our goal is to articulate the underlying values of the GDPR and use them to formulate foundational principles for RS design. These principles are valuable because they have withstood intense philosophical scrutiny over centuries. The rights given to data subjects under the GDPR reflect a certain European understanding of the human person and have evolved over time. Ultimately, from the European perspective, data protection and privacy are tools aimed at preserving human dignity \cite[pg. 89]{lynskey2015foundations, floridi2016human}. 

Accordingly, two crucial legal notions must be introduced: \emph{informational self-determination} and its predecessor, the \emph{right to the free development of one's personality}. Informational self-determination is defined as ``an individual's control over the data and information produced about him,'' and is a necessary precondition for any kind of human self-determination \cite{rouvroy2009right}. In turn, self-determination is a precondition for ``a free democratic society based on its citizens' capacity to act and to cooperate'' \cite{rouvroy2009right}.

It is worth quoting the 1983 German Federal Constitutional Court's decision to get a feel for its perspective \cite{rouvroy2009right}:
\begin{quote}
    The standard to be applied is the general right to the free development of one's personality. The value and dignity of the person based on free self-determination as a member of the society is the focal point of the order established by the Basic Law (Grundgesetz). The general personality right as laid down in Art. 2 (1) and Art. 1 (2) GG serves to protect these values-apart from other more specific guarantees of freedom-and gains in importance if one bears in mind modern developments with attendant dangers to the Human personality. 
\end{quote}

In sum, the right to informational self-determination derives from a more basic and older right to the free development of one's personality, found in the German Basic Law. Self-determination is necessary to uphold the dignity of the human person in society and is inherently tied to one's capacity for political participation. This suggests that participatory design of recommender systems could indirectly play a role in fostering political participation in the public sphere. 

We note that the European concept of freedom is not an ``anarchic'' freedom that conceives of the person as distinct and separable from society. Rather, it is broadly in line with more modern social constructionist thinking that views language as a shared social activity.  Language and knowledge are not something one possesses in one's head, but ``something people do''--as a ``sociorational process'' deriving from a communal ``negotiated intelligibility'' \cite{gergen1992social}. We will return to these ideas when looking at Hegel's thought. 

\subsection{Habermas and Political Participation}
 To really understand the GDPR and its view of the person, we must view it within the broader context of the project of EU integration. The EU must somehow unite many individual, self-determining member states, each with its own history, culture, and language(s) under a common EU flag. The European project is about how to maintain national identity in the face of political integration and homogenization \cite{habermas2009europe}. In other words, how can the EU construct a supranational identity while respecting the autonomy and self-determined identity of member states? \footnote{In the case of the GDPR, individual member states have derogations (exceptions) to spell out the details for their particular national implementations of the EU-wide Regulation.}

The philosopher and social theorist J\"urgen Habermas provides an influential model to understand the European project and the role of data protection in political participation. Building on the Kantian/Enlightenment devotion to reason, Habermas views the democratic process as fundamentally based on communication and interaction with others in a public sphere of argument and debate \cite{habermas2009europe}. Habermas rejects the metaphysical idea of an abstract, ahistorical, and universal ``truth,'' replacing it instead with broad consensus and intersubjective agreement arising from the application of the process of \emph{communicative reason}. 

Truth claims are valid to the extent to which they were generated by a process of deliberative argument with others, in good faith, and according to fundamental principles (pragmatics) of discourse.  Instead of asking what a moral agent could will--while avoiding self-contradiction--as a universal maxim for all, Habermas asks what ``norms or institutions would a communication community agree to as representing their common interests after engaging in a special kind of communication or conversation?'' \cite[pg. 24]{benhabib1992situating}. 


For Habermas, the role of the media is crucial in developing informed persons who can function in the public sphere. Yet the mass media wields disproportional power to influence the opinions and ideas of private persons, limiting their ability to engage in deliberative politics \cite{habermas2009europe}. Perhaps most importantly, mass media can lead to passive a citizenry with little experience in face to face political dialogue and its norms of reciprocity and argument. Further, mass media have the power to influence public opinion through the framing and selection of certain views and agendas at the expense of others. 

Consequently, we can understand data protection regulation as one antidote to the power asymmetries wielded by BBD platforms. Persons should participate in the project of deliberating upon and developing their own unique identities based on the personal data they generate in apps and on devices. As the legal scholar \cite[pg. 12]{lynskey2015foundations} notes, data processing ``exacerbates the information and power asymmetries between individuals and those responsible for personal data processing.'' In short, we see the GDPR as providing the data subject with the legal rights necessary to be an informed participant in one's self-determination and identity formation. But how should we understand the notion of identity as it relates to BBD?

\subsection{Humanistic Themes of the GDPR}
We claim the GDPR echoes themes from both the European Enlightenment and counter-Enlightenment. On the one hand, it stresses the \emph{conscious, reflective}, and \emph{rational} aspects of identity construction. On the other, it emphasizes \emph{individual expression} and rejects the goals of scientific \emph{absolutism}, \emph{reductionism}, and \emph{determinism} \cite{berlin2013roots}. As such, it aims to provide the minimally-necessary conditions for the exercise of unique human capacities. These capacities, identified and examined by key European philosophers, lay out the basis for these rights. 

We turn to philosophy because the grounds of human rights ultimately rest on moral considerations, not political ones. On this view, the ``fundamental conditions for pursuing a good life are various goods, capacities, and options that human beings qua human beings need, whatever else they (qua individuals) might need, in order to pursue the basic activities'' \cite[pg. 82]{cruft2015philosophical}.  As stated above, this includes self-development and an ability to participate and collectively deliberate in the political arena. 

\subsubsection*{Explicit Consent, Article 7}
The GDPR's focus on explicit consent for data collection and processing is one way in which conscious, reflective aspects of human experience are valued. For instance, the European Data Protection Board (EPDB) recently clarified its definition of \emph{clear and unambiguous consent to processing}\footnote{Accessed 30 June 2020, available at \\ \url{https://edpb.europa.eu/sites/edpb/files/files/file1/edpb_guidelines_202005_consent_en.pdf}}: 
\begin{quote}
    Based on recital 32, actions such as scrolling or swiping through a webpage or similar user activity will not under any circumstances satisfy the requirement of a clear and affirmative action: such actions may be difficult to distinguish from other activity or interaction by a user and therefore determining that an unambiguous consent has been obtained will also not be possible.
\end{quote}

\subsubsection*{Right to Human Intervention (''Human in the Loop''), Article 22}
Besides allowing data subjects to opt-out of automated processing, Article 22 has two key provisions: 1) data subjects have a right to obtain human intervention; 2) data subjects can contest the automated decision (and can also access the personal data used to make the decision) \cite{zarsky2016incompatible}. Additionally, data subjects must be ``informed'' about any automated-processes and provided with ``meaningful information'' about the logic of the decisions and the possible consequences of such automated-profiling (see, e.g., Recital 71). Such ``meaningful information'' also includes the ability to ``obtain an explanation of the decision reached'' (Recital 71). 

Recommendations, particularly in morally-salient contexts, such as dating or job hunting, potentially fall under this provision. Without knowing that such profiling--even with a ``human in the loop''--is occurring, and without understanding how the profiling was done, data subjects' rights to due process\footnote{Due process is a foundational principle of modern legal systems guaranteeing that procedures of the law are fairly applied to individuals. In the case of automated profiling, due process means notification that one is being profiled and the existence of some procedure through which one can contest the results of the profiling.} may be undermined. 

The GDPR's treatment of automated profiling (personalization in our case) belies an underlying distrust in the notion of quantifying the human experience. This is a theme explicitly tackled by German philosophers of the \emph{Sturm und Drang} movement and Romanticism more generally. Herder, for instance, eschewed the Enlightenment tendency to fit the particular under the general pattern, to quantify what he believed was inherently qualitative \cite{berlin2013roots}. For him, truth and goodness were not ideal, static Platonic forms, but relative to individuals residing in cultures with unique histories. If competing goals and values could not be resolved under a general algorithm, as Romanticists such as Herder believed, then keeping a human in the loop is one way to deal with the inevitable problem of value conflict in the ethical realm. The Kierkegaardian conflict of subsuming the infinitely complex individual under an abstract universal rule also can be seen here \cite{barrett1958irrational}. We detail more of these ideas in section 3.4. 

\subsubsection*{The Right to Be Forgotten, Article 17}
The genealogy of such a law traces back to the French \emph{le droit \`a l'oubli} (the ``right of oblivion'') which allowed convicted criminals who had served their time and had been ``rehabilitated'' to object to the publication of certain facts about their imprisonment \cite{rosen2011right}.  Viviane Reding, then Vice-President of the European Commission, claimed three goals for the right: strengthening individuals' rights to data protection, fostering an EU-wide ``Digital Single Market,'' and imposing greater responsibility on data controllers.\footnote{Accessed August 2nd, 2020, available at \\ \url{https://ec.europa.eu/commission/presscorner/detail/en/SPEECH_10_700}} According to Reding, it is ``important to empower EU citizens, particularly teenagers, to be in control of their own identity online.'' \footnote{Accessed August 2nd, 2020, available at \\ \url{https://ec.europa.eu/commission/presscorner/detail/en/SPEECH_12_26}}\footnote{\cite{marcia1966development} finds that young adults may experience any of several \emph{identity statuses} when exploring identity options and committing to identity goals. For our purposes, the statuses of \emph{foreclosure} and \emph{diffusion} are most relevant. Young adults in foreclosure have never fully ``explored and questioned the goals and values that were available'' to them. Further, diffusion is when young adults fail to make any commitments: ``they do not know what they want (or value) in adult life, and they are not, at the moment, looking to know'' \cite[pg. 193]{mcadams2006new}.} One question is whether this right applies to, for example, CF-based RS trained on the data of users who have exercised their right to be forgotten. 

The \emph{right to be forgotten} also mirrors the natural workings of autobiographical memory, in which the act of \emph{forgetting} is essential in constructing a self-narrative over time \cite{conway2000construction}. Narratives are the outcome of a continual cognitive process by which human experience is shaped into ``temporally meaningful episodes'' \cite{polkinghorne1988narrative}. Persons actively take part in constructing self-narratives to understand themselves, their behavior, and their roles in society \cite{mcadams1996personality}. The ability to reconstruct one's personal narrative is particularly important for young people \cite{marcia1966development}. We explore narrative identity more deeply in section 4.

\subsubsection*{Rights to Access and Modify Personal Data, Article 12}
Against scientific determinism, the GDPR upholds individual subjectivity and expression in creating one's digital representation \cite{rouvroy2009right}, where the very \emph{choice} of exercising one's rights inserts noise into the predictive signal of our in-app and on-device behaviors. Data subjects' rights to access, delete, and modify their personal data grant them the ability to both more deeply understand themselves and subjectively narrate their personal identities over time. Though the technological means for doing so are currently limited, data subjects can already modify their names, their genders, and drop nationalities \cite{andrade2012oblivion} . 

The right to modify one's personal data to fit one's life narrative provides an expression of \emph{agency} in that data subjects could potentially \emph{choose} the description under which their behaviors are understood, thus solving what Pariser calls the ``one identity problem'' of personalization \cite[pg. 65]{pariser2011filter}. The concept of reificiation is particularly applicable here. Reification is the process through which people believe the contingent manifestations of ideology are understood as facts of reality \cite{pitkin1987rethinking}. As the sociologists Berger and Luckmann write, ``Reification is the failure to recognize that [social order] as humanly engendered. The decisive question is whether one still retains awareness that... the social world was made by men--and therefore, can be remade by them''\cite{pitkin1987rethinking}. Put simply, the fear is that, as BBD platforms increasingly collect our personal data, their ``thin'' versions of our digital identities may begin to replace our ``thick'' personal identities. In the extreme, these thin digital identities may then be confused with our ``true'' ones, to our own detriment.

As the process of reificiation shows, without the GDPR's rights to access and modify their personal data, data subjects may fail to realize the extent to which their identities and behaviors have been classified according to the arbitrary rules of the BBD platform. RS designers, in order to avoid this problem of ``presumptuousness'' \cite{king2020presumptuous}, could provide users with notifications to confirm their intentions and goals. For example, one step towards this would be to ask data subjects, ``You did XYZ, but did you mean ABC?'' Doing so would give them the agency to decide the description under which their behaviors should be understood.

\subsubsection*{Right to Data Portability, Article 20}
Further, rights to access and download one's personal data in portable, machine-readable formats permit the \emph{reflective endorsement} of in-app and on-device behaviors. Coupled with open source data analytics platforms, data subjects could potentially analyze their own personal data to learn more about their captured behaviors on BBD platform. This is ostensibly the goal of the ``quantified self'' movement \cite{lupton2016quantified}. Downloading and then analyzing their own data allows data subjects to put a mirror to themselves in a new way. Even if they discover that their captured data bears no relation to their self-narratives, they have seen themselves in a new light. Moreover, the \emph{dignity}--meaning the worth or fittingness--of the human person, a crucial notion in the GDPR \cite{floridi2016human}, is expressed in the focus on reflective endorsement of behaviors and habits. The GDPR affirms the Socratic belief that a life without this self-reflective capacity is not properly a \emph{human} life.


\label{keythinkers}
\subsection{Key Thinkers of the Enlightenment and the Romantic Movement}
To illustrate some of the major threads from which we base our interpretation of the GDPR, we will highlight the ideas of just a few of the many Enlightenment and Romantic philosophers. We briefly highlight ideas from Kant, Hegel, Hamann, Herder, and Kierkegaard to bring out these themes. 

\subsubsection*{Immanuel Kant}
Kant's attitudes towards autonomy, self-determination, and the public use of reason are best illustrated by a short passage from his 1784 essay, \emph{Answering the Question: What Is Enlightenment?}\footnote{The German \emph{Unm\"undigkeit} is here translated as \emph{minor}, but has the sense of the English \emph{immature} or \emph{dependent.}} 

\begin{quote}
    Laziness and cowardice are the reasons why such a large part of mankind gladly remain minors all their lives, long after nature has freed them from external guidance. They are the reasons why it is so easy for others to set themselves up as guardians. It is so comfortable to be a minor. If I have a book that thinks for me, a pastor who acts as my conscience, a physician who prescribes my diet, and so on--then I have no need to exert myself. I have no need to think, if only I can pay; others will take care of that disagreeable business for me. 
\end{quote}

This excerpt reveals much about the European legal conception of autonomy and self-determination through one's critical use of reason. Our dignity rests on our capacity for reason and is the normative foundation for securing various data protection and human rights. To fail to exercise these rights is an affront to our dignity as rational beings. We are obligated to use them to progress in our self-understanding, to break free from our self-imposed immaturity. As rational beings, we possess a potential much greater than our habits, instincts and impulses. Realizing this potential, however, requires courage. For Kant, passivity is the mark of immaturity--thus we see Habermas emphasizing the process of communicative \emph{action} and critical debate in the public sphere. 

In the modern context of RS, we see Kant's disdain for the blind acceptance of recommendations without critical reflection of whether they might be reflectively endorsed by us \cite{frankfurt1988freedom}. We must \emph{dare to know} how these recommendations were made and understand the kinds of personal data involved in generating them. For instance, to what extent are they based on the predictive signals of our nonconscious goal directed behaviors? What about when our in-app behaviors are the result of dark patterns and hypernudges? Kant would implore us to take an active role in the public debate around the regulation of our personal data. 

Besides his important contribution to Enlightenment thinking, Kant's metaphysical conclusions famously resulted in a distinction of human knowledge into two domains: the intelligible (things in themselves) and the sensible (things as they appear to us). Kant concedes human knowledge has its limits, and we cannot know things in themselves, though we may employ reason to speculate on their nature.  Insofar as we possess reason and act ``in accordance with maxims of freedom as if they were laws of nature,'' we belong to a ``kingdom of ends,'' although our physical bodies firmly reside in the sensible, predictable realm of the natural world \cite[pg. 130-131]{paton1948moral}. For Kant, humans differ from other animals in this capacity to choose right or wrong (to follow the categorical imperative\footnote{The categorical imperative roughly states: one should only act according to a rule which one would also will to be a universal rule for all others in any particular circumstance.)}), and this capacity grounds our ascription of moral desert.

%

Later influential ``neo-Kantians,'' such as Max Weber and Wilhelm Dilthey, followed Kant's split between lived mental experience and our corporeality and aimed to delineate the ``human sciences'' from the natural sciences (i.e., \emph{Geisteswissenschaften} and \emph{Naturwissenschaften}). Dilthey writes, ``The hallmark of inner perception consists in the reference of a fact to the self; the hallmark of outer perception in the reference to the world'' \cite[pg. 36]{dilthey1989introduction}. The human sciences were different in that they allowed for not just explanations of events, but for ``understanding'' (\emph{Verstehen}) the phenomenological experience of persons \cite{dilthey1989introduction}. This ontological difference requires different epistemological methods of inquiry. For Dilthey, the method was ``epistemological self-reflection'' on moral facts, social relations, consciousness, and freedom. These \emph{kinds} of things could not be studied using methods from the natural sciences, which aimed to formulate universal, atemporal laws of nature. But Dilthey argues socio-historical reality cannot be reduced to a single principle or formula. Similarly, the human sciences (politics, anthropology, sociology, etc.) could not be founded on the approach based on the mechanical, mathematical model that Laplace used to predict the behavior of astronomical bodies \cite[pg. 39]{manicas2006realist}. 

\subsubsection*{GWF Hegel}
Hegel's philosophical project is notoriously complicated, but for our purposes we only consider a few important ideas in their more modern form. Hegel famously explained the human need for self-recognition in his master-slave dialectic. For Hegel, self-consciousness arises from the recognition of oneself by ``others,'' who are also self-conscious agents engaging in a kind of mutual recognition. The ``slave'' is marginalized in the sense that his ``self-consciousness is incomplete since he cannot find a reflection of his own autonomy and personhood in the master's eyes'' \cite[pg. 57]{chazan2002moral}. In other words, self-realization--recognizing oneself as  ``self''--fundamentally depends on the social recognition of other autonomous objects (persons in society) who recognize your existence as an individuated person with unique desires and goals. Our personal identities depend on this. If we are to see ourselves as independent and autonomous agents, we are required to see others this way too \cite[pg. 57]{chazan2002moral}. Hegel did not see us as disembodied Cartesian egos, but as fundamentally interdependent and social creatures embedded in social environments. 

Hegel represented a major advance from Kant in that he rejected the idea of an ahistorical, abstract, static, and universal faculty of ``reason'' separate from historical practices and cultural attitudes \cite[pg. 62]{fleischacker2013enlightenment}. In its place, Hegel inserted a teleology, or final goal, to reason towards which it inexorably moved in its quest for self-realization \cite{houlgate2006opening}. Hegel's addition of a temporal, historical aspect to logic means that something may be both A and not-A, perhaps at a different time. This idea will influence thinking about narrative identity over time. 

 Recently, social theorist and philosopher Axel Honneth updated these Hegelian ideas to include psychological notions of self-confidence, self-esteem, and self-respect in identity formation \cite{honneth2004recognition}. The Hegelian twist is that these can only be acquired through intersubjective, social experience with others. Honneth believes that justice and mutual recognition of personal identity are intrinsically related: ``The justice or wellbeing of a society is measured according to the degree of its ability to secure conditions of mutual recognition in which personal identity formation, and hence individual self-realization, can proceed sufficiently well'' \cite{honneth2004recognition}. Hegel's ideas about identity, dialectic, and mutual recognition have been developed and used by prominent gender theorists, such as Judith Butler, and various social movements for marginalized social groups, which are founded on claims of identity and autonomy \cite[pg. 9]{taylor2018remarkable}.

\subsubsection*{JG Hamann and JG Herder}
Hamann and Herder were the quintessential anti-Enlightenment thinkers who opposed Kant's glorification of universal, abstract reason. Hamann influenced the younger Herder and saw himself fighting against the increasing quantification and mechanization of life after Newton. Hamann's attack on Kant and other champions of reason could today be just as easily  directed at the data scientist. Enlightenment thinking enables the treatment of men as machines. Science, Hamman claims, was once an expression of our finite and creative capacity, but has now become a ``dictator which determines [our] position, morally, politically and personally'' \cite[Ch. 5, Par. 20]{berlin2013three}.  

Herder is arguably the first to give voice to the modern notions of authenticity, personal expression, belonging, and incommensurable ideals \cite[Ch. 3, Par. 26] {berlin2013roots}. When we speak of ``authenticity,'' or ``living one's truth,''  we draw on Herder's thought. For Herder, each of us has an original way of being human, making our quest for authenticity and self-realization unique \cite[Ch. 3, Par. 9]{taylor1991malaise}. Normalization should happen within, not between persons. Art is really the expression of a person's unique voice, attitude, and way of life of its author. Consequently, the object of art is inseparable from the identity of the artist; the subject could not be arbitrarily chopped off from the object without also destroying the identity of the object. Moreover, Herder was one of the first to speak of a shared language and soil as a bond between persons \cite[Ch. 3, Par. 31]{berlin2013roots}. Group identity and belonging confers a recognizable pattern to the activities of its members, giving rise to distinctive cultures, such as ``German'' or ``Chinese'' ways of doing things. There is no ``universal'' language or identity as human understanding is rooted in the shared symbols and languages of this root society. 

Lastly, Herder rejects the Enlightenment desire to find determinate answers to all questions by the application of a single method whose results might be reduced to a set of mutually derivable propositions \cite{berlin2013roots}.  Given our natural penchant for expression and belonging, Herder urged us to develop what we are to the fullest extent we can, to articulate our particular views in the richest way. Herder's praise of diversity was the apparent antidote to Enlightenment unity.  Today we see his ideas taking resurfacing in identity politics, postmodern relativism, and nationalism. 

\subsubsection*{S\o ren Kierkegaard}
Kierkegaard's ideas can be simply summarized as the primacy of the subjective over objective. Kierkegaard's legacy is the way in which he contrasted the Enlightenment universalism of Kantian ethics (his so-called categorical imperative) with the uniqueness and ``singleness of the single one'' \cite[Ch. 7, Sec. 3, Par. 6]{barrett1958irrational}. Kierkegaard and others after him rejected the idea that the particularities of individual situations should be abstracted away from in order to reach universal rules of conduct all reason-possessing persons could re-derive. Instead, the particularities of the individual situation are primary: (lived, embodied, fleshly) existence precedes essence (abstract, faceless, unitary).  He thus rejected the Kantian distinction of an unknowable abstract metaphysical realm from an observable, predictable, realm of experience. Kierkegaard conceived of human existence inherently textured by self-defining choices which are ``too dense, rich, and concrete'' to be represented as atomistic concepts \cite[Ch. 7, Sec. 2, Par. 9]{barrett1958irrational}. The human experience cannot be abstracted into a concept, but must be lived, experienced, and embodied as a subject of life. 

\subsection{Philosophical Lessons for RS Design}
From Kant's moral philosophy, we might claim that nonconscious behaviors are outside the realm of the moral because they were not the result of applying a universal maxim. This means that assertions about our logged behaviors reflecting our interests and preferences are not truly \emph{moral} statements, which seems contradictory. After all, we might define our interests as preferences precisely as those things which we believe to be good. In any case, a Kantian might conclude that if the logged behavior used to make predictions and recommendations stems from nonconscious sources, these should not be taken to represent our moral identities.

From the Romanticists Hamann and Herder, we might claim that that data scientists at BBD platforms create ``arbitrary divisions of reality'' (i.e., selectively log behaviors in apps and devices) in order to build ``castles in the air'' (predictive models of one's ``true'' preferences) \cite{berlin2013roots}. Put into today's language,  pre-defined categories of recorded behavior, which are in turn used by RS to make predictions, are categories of the data scientist's choosing and do not exhaust the reality of the data subject as he interacts with a device or app. Even in so-called ``context-aware'' recommender systems \cite{adomavicius2011context}, interpretations of the various contexts are one-sided and fail to account for contested interpretations by the data subject. This problem will become worse if BBD platforms continue moving towards reinforcement learning-based RS. 

From Hegel's story of the master-slave relationship we can derive an important principle for participatory RS design. We claim the GDPR's right of access and modification can support this dual-process of identity formation, particularly when combined with a narrative approach that evolves over time. Much as the master-slave dialectic shows, downloading and viewing one's logged behaviors used by an RS permits one to see oneself from the gaze of the system designer and potentially choose to change one's self-understanding as a result. At the same time, the act of negotiating meaning on the part of RS designers opens the possibility of enlarging their own self-understanding as a result of understanding the differing perspectives of others.

From Kierkegaard, we could view the GDPR and its rights that allow one to obtain, modify, and delete one's data potentially confront one with a series of self-defining choices.  At the same time, Kierkegaard would object to most RS ``personalization'' that relies on optimization of a global objective function on aggregate users' data, since the parameters of such a function are not strictly related to \emph{my} personal data alone. Similarly, Kierkegaard might argue that CF and its reliance on the logged behaviors of \emph{others} is irrelevant for \emph{my particular choice}. 

\label{narrativesec}
\section{Narrativity and Narrative Identity}
The story of human communication properly begins in \emph{logos}, or discourse. Logos subsumed the concept of everything from word, thought, narrative, story, to reason and rationale \cite{fisher1985narrative}. Only later, through a distinction made between \emph{mythos} and \emph{logos} by Plato and Aristotle, did logos become associated with a specific kind of philosophical and technical discourse, leaving us with the word \emph{logic} today. Mythos, on the other hand, became associated with poetry, while rhetoric uneasily fell somewhere in between \cite{fisher1985narrative}. 

The work of Jacques Derrida has notably questioned the assumptions behind this split, which has deeply influenced the course of Western metaphysics and mathematics. Other important thinkers, such as J\"urgen Habermas and Chaim Perelman \cite{perelman1982realm}, have tried to reclaim reason's original connection with rhetoric, communication, and dialog as a compromise position. 

Today, the modern concept of narrative plays an increasingly unifying role in a variety of disciplines, from clinical, developmental, social, and cognitive psychology to philosophy, literature, and even neuroscience \cite{singer2004narrative, damasio1999feeling}. The discussion here draws mostly from the philosophical perspective. Our claim is that narrative identity, with its dynamic and diachronic aspects, can unify both moral and social identity and possesses an explanatory force that accounts for human meaning and experience.

\subsubsection*{Self and Identity} 
According to identity theorists in sociology and psychology, the self is fluid and occupies multiple social roles or group identities that coexist and vary over time. The dynamism of the self-concept is reflected in its numerous representations. Persons have representations of a past, present, and future self; and also an ideal, ``ought,'' actual, possible, and undesired self \cite{markus1987dynamic}. Identity theorists highlight how a unified self-conception arises from the variety of meanings given to various social roles the self occupies \cite{stryker2000past}. 

\subsubsection*{Social and Moral Identity}
Common types of social identities relate to one's ethnicity, religion, political affiliation, job, and relationships. People may also identify strongly with their gender, sexual orientation, and various other ``stigmatized'' identities, such as being homeless, an alcoholic, or overweight \cite{deaux1991social}.\footnote{Most of these social identities would be \emph{sensitive personal data} under the GDPR, and they would be reviewable and modifiable by data subjects themselves.} Social psychologists generally agree our self-concept is one of the most important regulators of our behavior.

Psychologist and linguist Michael Tomasello contends that our membership in a linguistic community binds our social and moral identities. The reasons we give for our behaviors are related to our role and status within this community. Starting at a young age, children must make their own decisions about what to do and which moral and social identities to form \cite[pg. 288]{tomasello2019becoming}. Children make these decisions in ways justifiable both to others in their community and to themselves. The connection between the social and the moral lies in the way in which the reason-giving process to others was, in time, \emph{internalized} into a form of normative self-governance. Our psychical unity requires we do certain things in order to continue to be the persons we are, seen from both the inner perspective (self, private) and outer (other, public). 

Personality psychology has also begun to study the formation of moral identity. For example, \cite{aquino2002self} show how highly important moral identities--the collection of certain beliefs, attitudes, and behaviors relating to what is right or wrong--can provide a basis for the construction of one's ``self-definition.'' In short, our social and moral identities are crucial to both our self-understanding and our understanding of others. 

\subsubsection*{Narrative Identity}
Our real focus, however, is on narrative identity, as it encompasses both moral and social identity. Moral and social identities are synchronic (cross-sectional) structures.  They are how we represent ourselves to ourselves at particular points in time. But we have not yet explained how these identities evolve over time. For that, we need a diachronic (longitudinal) account of identity. 

The discussion below largely follows work by the influential psychologist Jerome Bruner and the philosopher Paul Ricoeur. According to Bruner, narratives ``[operate] as an instrument of mind in the construction of reality'' \cite{bruner1991narrative} and consist of many unique features including:

\begin{itemize}
\item \textbf{Diachronicity}: narratives account for sequences of ordered events over human time, not absolute ``clock time.''
\item \textbf{Particularity}: narratives are accounts of temporally-ordered events told from the particular embodiment of their narrator(s).
\item \textbf{Intentional state entailment}: within a narrative, reasons are intentional states (beliefs, desires, values, etc.) which act as causes and/or explanations.
\item \textbf{Hermeneutic composability}: gaps exist between the text and the meaning of the text. Meaning arises from understanding relations of parts to whole.
\item \textbf{Canonicity and breach}: narratives are more than just pointless ``scripts.'' They often arise in order to explain an anomalous event after the fact.
\item \textbf{Referentialility}: realism in narrative derives from consensus, not to correspondence to some ``true'' reality.
\item  \textbf{Genericness}: there are cultural patterns to various narratives, each expressing a version of human plights and tragedies.
\item  \textbf{Normativeness}: if narratives arise after breaches, this presupposes certain norms guiding our expectations about what will/should happen.
\item  \textbf{Context sensitivity and negotation}: readers of a text ``assimilate it on their own terms'' thereby changing themselves in the process. We negotiate meaning via dialogue.
\item  \textbf{Accrual}: an individual narrative can spread throughout a culture, thereby allowing it to grow and evolve collectively.
\end{itemize}

In contrast to Bruner's more psychological account, Paul Ricoeur combines an account of ethics with narrative identity. His goal is nothing less than to express a vision of the ``good life with and for others in just institutions'' \cite[pg. 240]{ricoeur1994oneself}. The philosophical question he poses is: how can one and the same person grow and change over time? How can such seemingly contrary concepts as identity and diversity be reconciled in one person? 

He answers via the notions of \emph{idem} and \emph{ipse} identity \cite{ricoeur1994oneself}. Idem is synonymous with sameness, typically associated with our character. Ipse is selfhood of self-constancy. It is what allows others to count on a person, and what makes him \emph{accountable} when making promises to others. Ricoeur's narrative account solves the problem of connecting permanence in time of character (sameness) with self-constancy (a kind of promise to vary within limits) \cite[pg. 166]{ricoeur1994oneself}.\footnote{Ricoeur states that self-constancy is the response of ``Here I am!'' when someone in need asks ``Where are you?'' \cite[pg. 165]{ricoeur1994oneself}.} Through the diachronicity of narrative, we unite our moral and social identities over time, giving rise to the uniqueness of persons. Good narratives and literature arise from imaginative variations between these two poles of sameness and selfhood over time. 

Ricoeur understands the self as fundamentally the ``character-narrator'' of its own history, ``emplotted'' between conflicting demands for concordance and a natural, entropic drive towards discordance \cite[pg. 141]{ricoeur1994oneself}.  Emplotment confers a unity, an internal structure and completeness to the story \cite[pg. 143]{ricoeur1994oneself}, giving one a coherent life story. However, if the discordance between events in the plot becomes too great, one's identity becomes threatened. If the discordance grows too wide, one's personal identity can dissolve in an ``identity crisis.'' Similarly, schizophrenia may be viewed as an example where the coherence of one's narrative identity has unraveled.

Similar to Aristotle's notion of \emph{mythos}, narrative events differ from mere occurrences by virtue of a certain ``effect of necessity'' or expectation from which they arise. Much as Bruner describes, only after the fact do events become part of a narrative and their configuration within this narrative--their relation as a part to the whole of the story--is what accounts for their meaning and explanatory force \cite[pg. 142]{ricoeur1994oneself}. We can only understand the conclusion of a narrative by reference to the earlier parts of the story which brought us there. As philosopher Charles Taylor puts it, ``what we grasp as an important truth through a story--be it that of our own life, or of some historical event--is so bound up with how we got there--which is what the story relates--that it can't simply be hived off, neglecting the chain of events which brought us here'' \cite[Pt. III, sec. 8]{taylor2016language}.  

Unlike scientific observations or chronicles of events in sequence, experiences recounted in narratives always have some moral tinge of approval or disapproval, rightness or wrongness. Relating to the hermeneutical issues of context, Ricoeur writes that imputability is the ``ascription of action to its agent, under the condition of ethical and moral predicates'' \cite[pg. 292] {ricoeur1994oneself}. Finally, in narrative the notion of absolute time is unnecessary because the character has the power to initiate a ``beginning of time,'' and assign a beginning, middle, and end to an action \cite[pg. 147]{ricoeur1994oneself}. 

\subsection{Narrative Accuracy and Epistemic Injustice}
In this section we now attempt to bring together narrative identity and narrative accuracy using \cite{fricker2007epistemic}'s concept of epistemic injustice. In philosophy, epistemology is the study of knowledge and its foundations. In order to apply her ideas to the realm of RS, we will, however, need to take some interpretive liberties. This discussion will ground the ethical vocabulary for thinking about and discussing the narrative accuracy of RS and their training data.

Typically injustice is understood as a kind of fair or fitting distribution of goods--one receives what one is properly owed. That is not quite what Fricker means by the term, as credibility is not really a finite good \cite[pg. 20]{fricker2007epistemic}. She is interested in injustice as it relates to disrespecting one in one's ``capacity as a knower.'' The astute reader will recognize the Kantian themes here: Fricker starts from the premise that we know ourselves best and have an equal stake in truth claims.  As \cite[pg. 8]{sherman2019overcoming} point out, epistemic injustice ``undermines individuals' trust in their own judgment and reasoning [and diminishes] their sense of intellectual agency.'' There is equally a Hegelian aspect to epistemic injustice in that it requires a mutual recognition of the perspective and experience of others, particularly those in positions of asymmetrical epistemic power (i.e., data subjects relative to data collectors). 

\subsubsection*{Testimonial and Hermeneutical Injustice} Fricker develops two forms of epistemic injustice applicable to the case of data subjects receiving personalized recommendations. First, \emph{testimonial injustice} might occur when prejudice or bias leads a data collector to give a ``deflated level of credibility'' to a data subject's interpretation of a recorded action or event, including a recommendation \cite[pg. 1]{fricker2007epistemic}. For example, if a data collector only uses nonconsciously-generated BBD and does not weight explicit feedback, a kind of testimonial injustice has occurred. Another example might be that a BBD platform allows users to ``downrate'' bad recommendations, but these are not actually factored into changing to the recommendations. From a Bayesian perspective, we can also conceive of testimonial injustice an example where uncertainty in model selection (ignoring the subjective ``model'' of the data subject) is ignored in favor of the pre-defined model of the RS designer. Even more generally, we can view it as examples of the problem of data fusion, or information quality, with ethical implications. 

In contrast, \emph{hermeneutical injustice} may arise when a data collector or data collection platform lacks the ``interpretive resources'' to make sense of the data subject's lived experience, thereby putting him at a disadvantage \cite[pg. 1]{fricker2007epistemic}. The fundamental question is,  \emph{what counts as what?} Currently the categories of events recorded by BBD platforms are typically pre-defined by system designers without any input from platform users, for instance. If designers of RS systems do not consider the diversity and richness of data subjects' intended actions, values, and goals while using the system, hermeneutical injustice will be unavoidable. Under one interpretation of an event, we may generate statistical regularities, while under another we may get different statistical regularities which become encoded in the parameters of ML models. It follows there is no one ``best'' representation or encoding of BBD. There are simply different representations under different interpretations about what counts as what. 

One way to potentially mitigate this would be for BBD platforms to ask users for explicit clarification of the meaning of key events and actions recorded within an app and used in predictive algorithms to generate recommendations. We saw earlier that RL systems are at particular risk for hermeneutical injustice towards their data subjects. 

With these ideas in mind, we can say the narrative accuracy of RS can be low through either weighing certain kinds of evidence too low, or when failing to negotiate the meaning of events or recommendations with data subjects. Fortunately, the GDPR gives data subjects some legal tools for engaging with and modifying their personal data generated and stored on BBD platforms. So if data subjects exercise their rights under the GDPR, and RS designers move away from BBD towards more explicit feedback, as in \cite{ekstrand2016behaviorism}, narrative accuracy can be shaped and improved in a communicative process with the data subject. 

This whole process of seeing oneself through the eyes of others (the BBD platform and the RS recommendations) leads to self-realization and the opportunity to reflectively endorse certain behaviors online under a more appropriate description, as seen from the perspective of the data subject. When data subjects can choose the description under which their actions take place and are recorded, it opens the possibility of understanding their experience online instead of forcing it to fit the designers' interpretation of ``what counts as what.'' This communicative process of narrative shaping through dialog with the data subject can be seen as one approach to participatory design of RS.

\subsection{The Dangers of Subjectivity and Narrative}
The downside to emphasizing the subjectivity of human experience in narrative form is that we might lose the possibility of a common, inter-subjective understanding of ``truth.'' This is a frequent criticism of postmodern thought. The thinking goes that we must therefore accept nihilism, or its equally-disagreeable twin, relativism. But doing so would be to open the door to difference and creativity, only to close it by being overwhelmed by what we find. 

Fortunately, many thinkers have already grappled with this problem. Although they agree there is no one single method that can deliver absolute knowledge, we can find minimal criteria of soundness that can provide a practical grounding for a theory of truth. For instance, we might try to articulate ``principles of charity'' \cite{davidson1973very} in interpreting and ``good faith'' in constructing narratives. We briefly sketch two attempts at grounding the truth claims of narrative knowing. 

The philosopher Marya Schechtman has described two ``constraints'' on the content of such personal narratives, if we are to realize our self-interests, get moral credit for our efforts, and take responsibility for our actions. These are the \emph{reality constraint} and the \emph{articulation constraint}. According to \cite[pg. 79]{atkins2010narrative} the reality constraint requires narratives to have some minimal level of coherence with reality. The articulation constraint says that the maker of a narrative must be capable of some minimal level of reflective articulation of one's actions and thoughts in order to properly be responsible for them. 

In communication studies, Walter Fisher elaborated the notion of narrative rationality and gave qualitative criteria for assessing it \cite{fisher1985narrative}. Fisher's criteria for assessing narratives break down into 1) narrative probability (coherence): does the story hang together and is it free of contradictions?; and 2) narrative fidelity (correspondence): are the reasons given logically sound and do they reflect a consistent set of values? \cite{warnick1987narrative}. 

These examples show that by embracing narrative forms of knowledge, we are not forced to abandon all criteria and any hope for consensus on ``truth.'' In fact, as the pragmatist philosopher Richard Rorty has famously argued, social convention, rather than representational similarity, can serve as the basis for a theory of truth \cite{rorty2009philosophy}. In Rorty's version of pragmatism, truth is defined relative to human goals and purposes which are articulated through a process of communication and social consensus over time. 

\section{Conclusion, Limitations, and Future Work}
Humanistic personalization aligns with the spirit of \emph{Human Centered AI} \cite{shneiderman2020human}. Making and sustaining a coherent digital self-narrative is a uniquely human capacity which we cannot leave up to others or outsource to automated agents. This sentiment is shared by the GDPR and IEEE EAD principles. We are characters in the stories we tell about ourselves. We know which events define us, we know which values drive us, we know the causes (reasons) behind our actions. And if we do not, we have the capacity to try to find out. 

The corporate owners of BBD collection platforms and RS designers may make claims to the contrary based on our observed behaviors, but we believe that rights to informational self-determination trump these assertions. Humanistic personalization promotes \emph{individual autonomy} \cite{o2002autonomy} by cultivating unique desires and personalities and by giving us the chance to create our own ``moral laws'' which guide us in our decisions and explain our actions.  

Ultimately, as postmodernists have pointed out, problems of ethics and interpretation are inseparable. What we believe to be true influences our decisions about what is right. But if meaning is socially constructed, data subjects alone cannot solve these problems. It will take both a community and good faith communication to work out the ``rules'' of our common language game. The designers of RS will need to play a larger role in this dialectic of meaning negotiation and identity formation in the digital sphere. After all, if the original meaning of \emph{category} is to ``publicly accuse,'' the data subject, as a member of the public, should play a part in that process.

A major limitation of our GDPR-based approach is that we assume our readers will agree with the European conception of the human person we have laid out. We assume readers share similar beliefs about what makes a human person unique and which capacities might be so essential as to require rights (and corresponding duties) to secure them. In an increasingly splintered yet interconnected digital world, searching for grand narratives of reason and truth appears misguided. Yet we believe the compromise position, based on narrative understanding and the dialogic participation of diverse groups, is the only way to avoid relativism and nihilism. 

Beyond personalization, a focus on narrative could have wide-ranging consequences for the future of AI/ML. If we are to ever ``crash the barrier of meaning in AI'' \cite{mitchell2019artificial}, we will need to crash through the barrier of narrative. Further, the inherent intelligibility of narrative could be useful in the emerging area of explainable AI, especially where regulations such as the GDPR give data subjects rights to clear, understandable explanations of algorithmic decisions. Lastly, due to its intuitive ``explanatory force'' \cite{velleman2003narrative}, narrative explanation could serve as a interesting lens for new approaches in causal modeling. How are events in narratives caused, and how can this be reconciled with claims of causation deriving from induction and observation?

Skeptics might counter that optimizing for narrative accuracy will require a trade-off in the ability of RS to accurately recommend items and predict specific behaviors--particularly non-conscious ones. Business profits may also be affected. Data scientists may need re-training. Nevertheless, the GDPR forces us to ask the question: Do we ultimately wish to represent ourselves according to the needs and interests of \emph{business} or \emph{humans}?

\bibliographystyle{acm}
\bibliography{Facct3pgUpdateSept4}

\end{document}